\documentclass[]{bmcart}

\usepackage{amsthm,amsmath}
\usepackage[utf8]{inputenc} %
\usepackage{setspace}
\usepackage{booktabs}
\RequirePackage{hyperref}
\usepackage{array}
\usepackage{graphics}

\usepackage[nameinlink,capitalise]{cleveref}
\crefname{section}{Sec.\!}{Secs.\!}
\Crefname{section}{Section}{Sections}
\crefname{figure}{Fig.\!}{Figs.\!}
\Crefname{figure}{Figure}{Figures}
\crefname{equation}{Eq.\!}{Eqs.\!}
\Crefname{equation}{Equation}{Equations}
\crefname{chapter}{Ch.\!}{Chs.\!}
\Crefname{chapter}{Chapter}{Chapters}

\def\includegraphics{}

\startlocaldefs
\newcommand\avg[1]{\left<#1\right>}

\newcommand\reponame{\href{https://github.com/network-cards/network-cards}{github.com/network-cards}}
\newcommand\tablefontsize{\small}
\endlocaldefs

\begin{document}

\begin{frontmatter}

\begin{fmbox}
\dochead{Research}

\title{Network Cards: concise, readable summaries of network data}

\author[
  addressref={aff1,aff2},                   %
  email={james.bagrow@uvm.edu}   %
]{\inits{J.}\fnm{James} \snm{Bagrow}}
\author[
  addressref={aff3,aff4},
  email={yyahn@iu.edu}
]{\inits{Y.-Y.}\fnm{Yong-Yeol} \snm{Ahn}}

\address[id=aff1]{%
  \orgdiv{Mathematics \& Statistics},             %
  \orgname{University of Vermont},          %
  \city{Burlington},                              %
  \state{VT},
  \cny{USA}                                    %
}
\address[id=aff2]{%
  \orgdiv{Vermont Complex Systems Center},
  \orgname{University of Vermont},
  \city{Burlington},
  \state{VT},
  \cny{USA}
}
\address[id=aff3]{%
  \orgdiv{Center for Complex Networks and Systems Research, Luddy School of Informatics, Computing, and Engineering},
  \orgname{Indiana University},
  \city{Bloomington},
  \state{IN},
  \cny{USA}
}
\address[id=aff4]{%
  \orgdiv{Network Science Institute},
  \orgname{Indiana University},
  \city{Bloomington},
  \state{IN},
  \cny{USA}
}

\end{fmbox}%

\begin{abstractbox}

\begin{abstract} %
The deluge of network datasets demands a standard way to effectively and succinctly summarize network datasets.
Building on similar efforts to standardize the documentation of models and datasets in machine learning, here we propose \emph{network cards}, short summaries of network datasets that can capture not only the basic statistics of the network but also information about the data construction process, provenance, ethical considerations, and other metadata.
In this paper, we lay out (1) the rationales and objectives for network cards, (2) key elements that should be included in network cards, and (3) example network cards to underscore their benefits across a variety of research domains. 
We also provide a schema, templates, and a software package for generating network cards. 
\end{abstract}


\end{abstractbox}

\end{frontmatter}

\section{Introduction}
\label{sec:introduction}

Network structure can be found in numerous complex systems and it provides a unifying framework to study those systems collectively~\cite{borner2007network,mitchell2009complexity,newman2018networks,menczer2020afirstcourse}.
Beyond academic interests, we also live in a connected world and any actions we take online leave digital traces that echo various socioeconomic networks~\cite{lazer2009computational}. 
Due to its broad appeal and usefulness, the network perspective is widely used across domains and \emph{network data} is ubiquitous in science and society.

Despite the universality and deluge of networks, there is currently no consensus nor standard procedures to report the characteristics of networks and their metadata. 
As argued in the case of documenting models and data in machine learning~\cite{mitchell2019model,gebru2021datasheets}, the lack of such standards---and the resulting lack of attention paid to important aspects of models and datasets---may lead to negative societal outcomes. 
This is equally true for network datasets, particularly given the ubiquity and relevance of network datasets across academia and industry. 

Here we introduce network cards\footnote{See also: \reponame{}.}, standard tabular summaries of network data that capture both metadata about the network and statistics describing the data themselves.
While we cannot expect a ``one-size-fits-all'' solution given the variety of networks that scientists consider, network card's ability to summarize the most basic network statistics will be broadly appealing and make network data more accessible. 
Network cards are intended to be flexible enough to describe a variety of rich network types such as multilayer and higher-order networks and can even extend to describe multiple networks simultaneously.
Network cards are also complementary to other efforts such as more detailed ``datasheets''~\cite{gebru2021datasheets}. 
Cards can provide a succinct summary of network-specific information which can be expanded upon when needed.
And, when aspects of a datasheet---which is focused on datasets for machine learning---are not exactly applicable, the network cards can be used to document both network statistics and key metadata that pertain to data provenance, ethics, privacy, and other concerns. 

A network card can provide researchers with a number of benefits.
Glanceable information about a network dataset allows researchers to quickly digest the most salient features of the dataset.
Network cards will answer the most basic questions about network data, such as: What are the nodes? What defines links? How big is the network? How dense? 
Awareness of the basic information provided by the network cards can prevent misinterpretation of network data and having a set of standardized statistics and information will encourage both data producers and users to pay attention to key details such as how and when the data were gathered or whether there are any important ethical considerations involving the dataset.

\section{The need for concise network summaries}
\label{sec:the-need-for-concise-network-summaries}

Most network researchers are familiar with the Zachary Karate Club~\cite{zachary1977} (\cref{tab:zachary_karate_club}), a very popular example network. 
But did you know that the original data contained eight different ``interaction contexts'' and can be considered a multiplex network?
This rich context is now mostly lost and rarely discussed because the most widely disseminated dataset for the Karate Club was the version where all contexts are collapsed down to binary edges. 

\begin{table}[t]
	\tablefontsize
    \centering
    %
\begin{tabular}{lp{7.25cm}}
\toprule
Name & Zachary Karate Club \\
Kind & Undirected, unweighted \\
Nodes are & Members of club at US university \\
Links are & Members consistently interacted outside club \\
Considerations & Heavily used as an example network \\
\midrule
Number of nodes & 34 \\
Number of links & 78 \\
Degree$^*$ & 4.588 [1, 17] \\
Clustering & 0.571 \\
Connected & Yes \\
Diameter & 5 \\
Assortativity (degree) & -0.476 \\
\midrule
Node metadata & None \\
Link metadata & None (original study included eight interaction contexts) \\
Date of creation & 1977 \\
Data generating process & Direct observation of club members during period 1970-72 \\
Ethics &  \\
Funding & None \\
Citation & Zachary (1977) \cite{zachary1977} \\
Access & \url{https://networkrepository.com/karate.php} (accessed 2022-02-12) \\
\bottomrule
\multicolumn{2}{l}{\footnotesize $^*$Distributions summarized with average [min, max].}
\end{tabular}
%
%

    \caption{Example network card for the Zachary Karate Club.
    A network card is a concise, three-panel, tabular summary of a network and associated information.
    The three panels summarize, from top, overall information about the network, the structure of the network such as its size and density, and meta-information such as where the data originated and any ethical considerations associated with the data.
    \label{tab:zachary_karate_club}}
\end{table}

As another example, consider protein-protein interaction (PPI) networks.
These data are collected through experimental assays that test whether proteins interact with one another. 
But not all assays are designed to detect dyadic (pairwise) interactions.
For example, whereas Yeast Two-Hybrid (Y2H) does test pair interactions in isolation~\cite{bruckner2009yeast}, Affinity Purification Mass Spectrometry (AP-MS) uses tagged bait-prey protein pairs to identify interacting clusters (complexes) of proteins~\cite{gingras2007analysis}.
In other words, the results of AP-MS assays will over-represent \emph{cliques}.
These different data generating processes have profound consequences for the final network structure, with AP-MS-derived networks exhibiting far more clustering than Y2H.
A researcher not recognizing these differences may draw inappropriate and biased conclusions, which may lead to potential harms down the line. 
This leads us to ask,
how to best retain critical information such as these experimental details, external to a network's structure, when disseminating the network data?

A contributing factor towards losing critical details about experiments and data over time may be information overload.
The scientific literature is estimated to double in size every 15--25 years~\cite{bornmann2015growth,fortunato2018science,bornmann2021growth}.
Furthermore, particularly in the case of network data, we are challenged not only by the growth of the absolute volume of papers, but also the breadth of the works that deal with network data.
Network science is a highly interdisciplinary field and researchers interested in network data come from all domains of research~\cite{borner2007network} and any efforts to retain critical experimental and data details should be both succinct---to mitigate information overload---and broadly accessible.

Alongside the literature's exponential growth,
many fields of research are in the midst of a replication crisis, where past work has been called into question~\cite{ioannidis2005why,collaboration2015estimating,nissen2016publication,cockburn2020threats}. 
Causes of this crisis include poor statistical practice~\cite{loken2017measurement,benjamin2018redefine,gosselin2020statistical} and poor data documentation~\cite{kanwal2017investigating,taylor2018crisis,rupprecht2020improving}.
Documenting the provenance of data is crucial for data-driven studies of networks, and there is a real need for a systematic, standard way to describe the various details of a network dataset, details that are not strictly part of the network topology itself and so are often lost as researchers share data files describing that topology but nothing else.

To retain critical details accompanying datasets in the face of information overload while accommodating broad interdisciplinary interest, we argue that standardization, portability, and succinctness of presentation are critical. 
Standardization is crucial not only because it acts as a shared \emph{checklist} that keep researchers from omitting important details, but also because it allows the development of shared understanding and tools. 
The more portable it is, the easier to prevent the loss of critical metadata. 
Succinctness of presentation is critical: a researcher should (correctly) understand the results of a scientific study as quickly (and accurately) as possible. 
As the writer of a scientific study, this can require hard choices when describing the results, using enough jargon for the intended audience but not more, enough technical detail for someone to replicate the study but not so much they cannot follow the results, and enough interpretation so the results are communicated and contextualized clearly but correctly.

Taken together, these factors---information overload, interdisciplinary network interest, and the need to document data---point towards the need for a succinct, standardized, broadly readable, and portable means to summarize network datasets. 
Our goal here is to propose a solution to meet these needs, the network card.

\section{Network cards}

Network cards were designed to achieve three properties:
\begin{enumerate}
    \item Concise. 
    A card should be compact and efficient, occupying one page at most. Concise presentation also helps the portability of the cards. 
    \item Readable. 
    Any researcher familiar with networks should instantly understand all card contents. 
    Cards should be as approachable to non-specialists as possible.
    \item General and flexible. 
    Works for all types of networks. 
    Can be adapted to special circumstances.
\end{enumerate}

Inspired by summaries of regression models, we propose a three-panel tabular layout for network cards, with the first panel providing overall information about the network, the second panel focusing on structural information related to the network's topology, and the third panel describing further meta-information such as availability of metadata, how the data were generated or gathered, and any ethical concerns to consider.

\Cref{tab:zachary_karate_club} shows a network card for the famous Zachary Karate Club, one of the prototypical example networks used in the literature.
From a glance at the card, a reader can deduce a number of salient details including where the network data came from, what constitutes nodes and links, the network's size, and where to go for more information.
Although the Karate Club is such a heavily used example, as we discussed above, less familiar may be some of the associated features of the club, such as when the data were gathered, and the fact that multiple interaction contexts were captured for the social ties in the network.
If this network card can ``follow'' the data and be readily available, it would be less likely that such information gets lost. 

We now discuss the contents of each panel of a network card in greater detail.
\Cref{sec:cardcontents} contains a complete description, derived from our schema, of all entries broken down by panel.

\subsection{Overall information}
\label{subsec:overall-information}

The top of each network card provides an overall description of the network, a name, whether the network is undirected or directed, whether there are link weights, and any further considerations worth bringing attention to.
Sometimes left unstated, but absolutely crucial, are explicit definitions for the nodes and links.
Almost always the first question a network scientist asks when a collaborator brings them unfamiliar network data is what constitutes the nodes and what relationship defines the links. 
We believe that providing a prominent and explicit venue for displaying these definitions is one of the most valuable aspects of network cards.

\subsection{Structure}
\label{subsec:structure}

The second panel of a network card provides basic summary statistics for describing the structure of the network. 
We intentionally rely on the most common and broadly understood network statistics, ensuring cards are readable to as many researchers as possible.
Our goal is to summarize the size, density, and connectivity of the network, so we report the number of nodes and links, the average degree, average clustering coefficient, whether the network is connected, and the degree assortativity of the network.
If the network is connected, we report the network's diameter. 
If it is not connected, we report the number of connected components, the proportion of nodes in the largest component, a summary of the distribution of nodes per component (component size), and the diameter of the largest connected component.
\Cref{tab:M_PL_058}, which we discuss in greater detail in \cref{sec:examples}, demonstrates these statistics.

\begin{table}[t]
	\tablefontsize
    \centering
    \centerline{
\begin{tabular}{lp{8cm}}
\toprule
Name & M\_PL\_058 \\
Kind & Undirected, unweighted \\
Nodes are & Plants and pollinators \\
Links are & Pollination interactions \\
Considerations & Bipartite [32 plants, 81 pollinators] \\
\midrule
Number of nodes & 113 \\
Number of links & 319 \\
Degree$^*$ & 5.646 [1, 28] \\
Clustering & 0 \\
Connected & 2 components [98.23\% in largest] \\
Component size & [111, 2] \\
Diameter & n/a \\
Largest component's diameter & 6 \\
Assortativity (degree) & -0.379 \\
\midrule
Node metadata & Species name \\
Link metadata & None \\
Date of creation & Spring 2005 \\
Data generating process & Field observation at study sites in Natural Park of Cap de Creus in Catalonia, Spain; data retrieved from web-of-life.es \\
Ethics & Work complied with the current laws of Spain \\
Funding & Integrated European Project Assessing Large Scale Risks to Biodiversity with Tested Methods, Ministerio de Ciencia y Tecnología projects Efecto de las Especies Invasoras en las Redes de Polinización, Determinantes Biológicos del Riesgo de Invasiones Vegetales \\
Citation & Bartomeus \emph{et al.}\ (2008) \cite{bartomeus2008} \\
Access & \url{https://www.web-of-life.es/networkjson.php?id=M_PL_058} (accessed 2022-03-10) \\
\bottomrule
\multicolumn{2}{l}{\footnotesize $^*$Distributions summarized with average [min, max].}
\end{tabular}
}
%
%

    \caption{Network card for a plant--pollinator network.
    This example shows how to highlight the details of a bipartite network.
    \label{tab:M_PL_058}}
\end{table}

Examining some network cards, one realizes there is some redundancy in these quantities. 
For example, the average degree $\avg{k}$ comes directly from the numbers of nodes $N$ and links $M$ (via $\avg{k} = 2M/N$), which we already report.
While it is thus not strictly necessary, we believe it is worth including an entry describing the degree because (1) we can summarize more of the distribution than its first moment (see our notes on statistical summaries, below), (2) it saves readers some (albeit simple) mental arithmetic while reading, and (3) it provides a nice way to compare networks: often, differences in average degree are more important than different numbers of nodes or links. 
The clustering coefficient is another case where its inclusion is useful even if it appears unnecessary: for a bipartite network, clustering is zero by definition. 
However, we feel it is worth always including to maintain consistency and in this case to reinforce as a ``sanity check'' this natural feature of a bipartite network. 
(Imagine the dismay of someone writing code to process their data and discovering triangles in what should be a bipartite network!)

\subsection{Meta-information}
\label{subsec:meta-information}

Beyond general information about the network and summaries of the network's topology, we sought to capture additional salient details that describe the network without directly relating to the network data itself.
The presence of metadata describing the elements of the network is top of mind, and we include entries for noting metadata describing nodes and describing links.
Examples of node metadata include 
demographic variables for individuals in a social                      network
or
gene ontology terms for proteins in a PPI network.
Examples of link metadata include
the mode of communication in a contact network (e.g., text or phone call in a mobile phone network),
and
the link type (or interaction context) in a multilayer or multiplex network.
Often these metadata draw on unique facets of the network under study and can enable novel research questions by drawing comparisons between features of the network and values in the metadata.

Next, we include information on the date of creation, allowing researchers to specify the time period of the dataset. 
In particular, we believe it is important to specify whether the network came from ongoing data collected over a long period.
One situation where this is useful is for comparing networks (see below): we can use the date entry to distinguish, for instance, a network measured in 2020--21 from that network measured during 2022--23.

An entry describing the data generating process is also included in the meta-information box. 
Broadly construed, this entry is intended to briefly describe how the network data were collected, measured, captured, or otherwise quantified.
A social network derived from mobile phone billing records can thus be distinguished from a social network derived from chest-worn proximity sensors and from a social network derived from manual observation by researchers.
All are ``social networks'', but these very different generating processes will have profound impacts on the form of the network and how the data should be interpreted.
Of course, networks are captured by all different manners of data generating processes, making it necessary to be flexible in what details to include, and some processes can be quite complex and difficult to describe succinctly.
Nevertheless, brief descriptions of the process, even relying on citations for more information, are invaluable and essential.

Lastly, and in many ways most importantly, we provide entries for reporting ethical concerns, funding disclosures, a citation, and an access field describing where the data were retrieved.
The ethics entry can be used highlight particular concerns such as whether the network data required informed consent or whether the network data can be freely shared or not. 
Researchers may also wish to indicate whether the data are inappropriate for certain applications.

\subsection{Important considerations}
\label{subsec:important-considerations}

Some important further considerations are worth noting.

\paragraph{Importance of meta-information}

While summary statistics of the topology can be extracted programmatically, data formats are often poor documents of ethical concerns, the meanings of nodes and links, the data generating process, and so forth, all of which are important for reproducibility and better understanding the data's nature.
Network cards are therefore a valuable additional document, providing a succinct record of these critical but often ephemeral information that are not readily captured in the network's topology.

\paragraph{Statistical summaries}

Most networks are large and distributions of numeric network properties are typically employed, perhaps the most popular being the network's degree distribution, the number of connections per node.
To include such information in a network card, we recommend using basic summary statistics, the simplicity of these statistics being key to our design goals. 
For most situations, a distribution can be summarized using the mean value along with a measure of range such as its minimum and maximum. 
However, many network statistics are heavy-tailed or broadly distributed and a more robust choice than the mean is the median.
Therefore, we propose basic summaries of the form, ``average [min, max]'' or ``median [5th percentile, 95th percentile]''.
One exception is if the quantity being summarized has only a few values. 
This can happen, for instance, when summarizing the component sizes of a network that is not connected but has, say, only two connected components---it does not make much sense to include summary statistics for only two values.
We recommend simply including a list of the observed values when there are five or fewer values; our implementation (see below) will do this automatically. 
We encourage users of network cards to always include a footnote describing exactly how distributions are summarized such as we use in \Cref{tab:zachary_karate_club}.

\paragraph{Absence of graphics}

Graph layout visualizations are often used to present networks and we considered including an entry for such a figure.
We decided against it for two reasons.
One, layouts are not always helpful or appropriate. They work best for sparser and smaller networks. Many networks are simply too large or dense to be readable in a two-dimensional projection, as anyone who has found themselves staring at a ``hairball'' graphic can attest.
Two, including graphics makes it more challenging to support formats such as plain text or spreadsheets. (This also rules out other graphical summaries such as histograms.) 
A purely written format makes such representations easy to produce.

\paragraph{Comparing multiple networks}

One particular strength of concise tabular summaries is that they are readily extended to multiple networks simply by adding additional columns. 
In other words, if a one-network card can be thought of as a two-column table, with the first column labeling each entry and the second for the entry's contents, then a ``multicard'' will have one column for labels and one column for each network.
A multicard can be useful in several scenarios, such as comparing different network extraction techniques for the same data,  capturing snapshots of a dynamic network, or even comparing replications of one study across different experiments. 
We show an example multicard in \cref{sec:examples}.

\paragraph{Special networks}

Networks are complex, and there is a whole zoology devoted to different forms of networks capturing all manner of different structural and dynamic properties.
Our standard network card accounts for all combinations of directed, undirected, weighted, and unweighted networks, but many other types of networks exist, some of which can also be handled well by our standard card but others may require further consideration.
We discuss ways to accommodate such networks.

Bipartite networks, where nodes form two disjoint sets and links exist only between nodes in different sets, can be accommodated with a standard card, but it is worth explicitly denoting the network's bipartiteness, as we do in \cref{tab:M_PL_058}.
Some statistics in the structure panel, such as the degree distribution, may be replicated to report the distribution for both node sets separately.

Temporal networks can be most naturally accommodated by either transforming them into a static network and noting its temporal nature as metadata (see \cref{tab:tnet_malawi_pilot}), comparing multiple snapshots of the network with a multicard, or adapting some entries of the card (particularly those in the structure panel) to describe the temporal network directly.
While this last option seems most appealing, it can in fact be the most challenging, as considerable research continues on how best to quantify temporal networks, and the results of this work have not yet experienced wide adoption compared to the more basic measures we used.
And in terms of describing the results of transforming the dynamic network into a static network, a network multicard with one column for each choice of transform can actually work very well to demonstrate and compare those transformations against one another.

Signed networks, where edges have positive and negative values associated with them, can be handled in a standard network card simply by denoting signedness in the considerations entry. This entry can also include a basic statistic for the overall proportion of negative links, for example: ``links are signed [23.5\% links are negative]''. 
For a signed network, it is crucial to denote the meaning of signedness in the ``Links are'' entry, for example: ``links are ally (positive weight) or adversary (negative weight) ties''.

Multilayer networks, where nodes are associated with one or more contexts or layers and links exist between or across layers, can also be handled by standard cards by denoting their layers as considerations and as node/link metadata. 
The structure of the different layers can be illustrated, at least when there are not too many layers, by either expanding each statistic, for example reporting layerwise the sizes, densities, and so forth, or by using a multicard where each column describes a layer of the network. 
It may be worth including multilayer-specific statistics in the structure panel of the card, but again relying on uncommon measures may not be worth the loss in readability to some audiences. 
Important considerations to note are if the network is multiplex, where nodes ``replicate'' across each layer and, closely related to multiplex, if the network is multirelation, where multiple links can exist between nodes (cf.\ \cref{tab:zachary_karate_club}). 
Both situations can be well described using the node and link metadata entries as well as the considerations entry as needed.

Lastly, higher-order networks or hypergraphs have recently seen increased interest.
A network card can immediately indicate whether the network is higher-order using the `Links are' entry, for example: ``Links are: social groups (hyperlinks)''.
Among the possible choices for additional statistics to describe the hypergraph's structure, we recommend at minimum adding a statistic for the distribution of link size (nodes per link). 
This basic quantity should be broadly understood by readers and captures much useful information.
Other statistics specific to higher-order networks may be worth including as well, again under the caveat that broad readability of the card should be maintained as much as possible.

\paragraph{Implementation}

To make network cards easy to use, we have created an open source implementation available online (\reponame{}). 
Our package facilitates generating network cards as tables and spreadsheets, and to read and write network cards in a standard format (we provide a schema). 
Currently, our package works with Python; over time, we hope to support other languages, such as R, MATLAB, and Julia.
We also provide templates for the common network types that researchers can complete manually.

\section{Examples of network cards}
\label{sec:examples}

We have assembled some example networks meant to highlight the usefulness of network cards as summaries across research problems involving network data.
These examples span social, biological and ecological research.

Our first example, already introduced in \cref{tab:zachary_karate_club}, is the famous Zachary Karate Club.
As noted earlier, it displays both commonly known properties of the network as well as hardly-discussed features (multiple interaction contexts). 

\Cref{tab:M_PL_058}, also previously seen, shows a network card for a bipartite plant--pollinator network~\cite{bartomeus2008}. 
This network, collected from field observations in Spain, highlights the card's flexibility at concisely capturing metadata.
In particular, it describes the bipartiteness including the numbers of plants and pollinators, the available metadata (species names are associated with each node), and the study details (when and where the data were collected). 
We also see the network is relatively dense, is not globally connected due to a single disjoint link, and the nodes are strongly degree dissortative.

Next, \cref{tab:tnet_malawi_pilot} shows the network card for a temporal contact network~\cite{ozella2021}. 
This network was captured from proximity sensors worn by participants.
We transform the temporal network data to a weighted network and the network card succinctly documents our processing.
As expected for most social networks due to triadic closure, this network is triangle-heavy, with a clustering coefficient greater than 0.5. 
Like the previous example, we again see that the network consists of two disconnected components, where one contains most nodes.
This network card also illustrates how one can describe ethical concerns for the study, including acquiring informed consent from study participants.

\begin{table}[t]
 	\tablefontsize
    \centerline{
    %
\begin{tabular}{lp{8cm}}
\toprule
Name & TNet Malawi Pilot \\
Kind & Undirected, weighted \\
Nodes are & Study participants \\
Links are & Close proximity interactions \\
Link weights are & Number of interactions \\
Considerations &  \\
\midrule
Number of nodes & 86 \\
Number of links & 347 \\
Degree$^*$ & 8.070 [1, 31] \\
Clustering & 0.527 \\
Connected & 2 components [97.67\% in largest] \\
Component size & [84, 2] \\
Diameter & n/a \\
Largest component's diameter & 5 \\
Assortativity (degree) & 0.0363 \\
\midrule
Node metadata & None \\
Link metadata & Time and duration of interaction \\
Date of creation & 2019-12-16 to 2020-01-10 \\
Data generating process & Study participants in a rural village in Malawi wore a low-power sensor on the chest to measure their proximity to other participants. Time of contact was recorded and transformed to link weights \\
Ethics & Written consent was obtained from all participants or their guardians (both, in the case of adolescents). Study approved by Ethical Committee at the University of Zurich (OEC IRB \#2018-046) and Ethical Committee at College of Medicine in Malawi (P.10/19/2825) \\
Funding & UNICEF Malawi and support from the Lagrange Project funded by the CRT Foundation \\
Citation & Ozella \emph{et al.}\ (2021) \cite{ozella2021} \\
Access & \url{http://www.sociopatterns.org/datasets/contact-patterns-in-a-village-in-rural-malawi/} (accessed 2022-03-24) \\
\bottomrule
\multicolumn{2}{l}{\footnotesize $^*$Distributions summarized with average [min, max].}
\end{tabular}
}
%

    \caption{Network card for a temporal contact (close proximity) network.
    This example highlights using a card to document transformation of a dynamic network to a (weighted) static network as well as describing ethical concerns (informed consent).
    \label{tab:tnet_malawi_pilot}}
\end{table}

\begin{table}[t]
	\tablefontsize
    \centering
    %
\begin{tabular}{lp{8cm}}
\toprule
Name & OpenFlights routes \\
Kind & Directed, weighted \\
Nodes are & Airports \\
Links are & Direct routes flown between airports (source node: departing airport, target node: arriving airport) \\
Link weights are & Number of routes \\
Considerations & Historical records, updated 2014 \\
\midrule
Number of nodes & 3425 \\
Number of links & 37595 (1 self-loop) \\
--- Bidirectional links & 48.8\% \\
Degree (in/out)$^*$ & 10.9766 [0, 238] \\
Degree$^+$ & 21.9533 [1, 477] \\
Clustering & 0.4692 \\
Connected & Disconnected \\
Assortativity (degree) & -0.0104 \\
\midrule
Node metadata & IATA airport codes \\
Link metadata & None (airline IDs, codeshare status, equipment IDs available in original data) \\
Date of creation & 2014 \\
Data generating process & Open flights data retrieved, codeshare routes removed, routes grouped by airport codes to get directed links and weights \\
Ethics &  \\
Funding & None \\
Citation & None \\
Access & \url{https://openflights.org/data.html} (accessed 2022-09-26) \\
\bottomrule
\multicolumn{2}{l}{\footnotesize $^*$Distributions summarized with average [min, max].}\\
\multicolumn{2}{l}{\footnotesize $^+$Undirected.}
\end{tabular}
%
%

    \caption{Network card for a directed network.
    This example highlights structural fields specific for directed networks. For directed networks we recommend explicitly describing the directionality of links by referring to source and target nodes.
    \label{tab:openflights_directed}}
\end{table}

As another useful example, in \cref{tab:openflights_directed} we present a network card for a directed network, in this case the network of direct flight routes flown between airports. 
Here, nodes are airports, represented with IATA airport codes, and links are directed, weighted links counting the number of direct flights between pairs of airports. 
Structurally, we report the number of bidirectional links as a proportion of all links, as well as summaries of the in-degrees, out-degrees and the degree treating the network as undirected. 
As we show in this example, for directed networks, we recommend always explicitly defining link directionality by referring to the source and target nodes in the ``Links are'' entry.

\Cref{tab:huri} meanwhile, shows the network card for the recently released HuRI, the human reference interactome~\cite{luck2020}.
This example shows how biological information can be put into a card, describing the gene metadata associated with nodes in the network and a brief description of the high-throughput assays used to infer protein-protein interactions (PPIs).
A researcher interested in these data will immediately know where they can turn to enrich their study with node metadata, in this case using standard GENCODE gene annotations.
HuRI also exhibits self-loops, capturing a small set of self-interacting proteins, and the card naturally draws attention to this information.

\begin{table}[t]
	\tablefontsize
    \centerline{
    %
\begin{tabular}{lp{8cm}}
\toprule
Name & HuRI \\
Kind & Undirected, unweighted \\
Nodes are & Human proteins \\
Links are & Binary protein interactions \\
Considerations &  \\
\midrule
Number of nodes & 8\,272 \\
Number of links & 52\,548 [480 self-loops] \\
Degree$^*$ & 12.705 [1, 500] \\
Clustering & 0.0592 \\
Connected & 72 components [98.51\% in largest] \\
Component size$^*$ & 114.889 [1, 8\,149] \\
Diameter & n/a \\
Largest component's diameter & 12 \\
Assortativity (degree) & -0.115 \\
\midrule
Node metadata & GENCODE v27 gene annotations \\
Link metadata & None \\
Date of creation & 2019 \\
Data generating process & Links inferred using a high-throughput three-panel yeast two-hybrid assay applied to pairs of protein-encoding genes taken from human ORFeome v9.1 \\
Funding & National Institutes of Health and others \\
Citation & Luck \emph{et al.}\ (2020) \cite{luck2020} \\
Access & \url{http://www.interactome-atlas.org/download} (accessed 2022-04-01) \\
\bottomrule
\multicolumn{2}{l}{\footnotesize $^*$Distributions summarized with average [min, max].}
\end{tabular}}
%
%

    \caption{Network card for a protein-protein interaction network.
    \label{tab:huri}}
\end{table}

Luck \emph{et al.}\ contrast HuRI with pre-existing PPI networks, one (``Lit-BM'') extracted binary interactions from literature curated datasets, another (``HI-union'') combined all previous screening experiments conducted by the research group with HuRI.
We compare these networks with HuRI in a multicard shown in \cref{tab:huri-multicard}.
With these summaries, we can succinctly define the similarities and differences in the networks, both in their data generating processes and in their structure.

\begin{table}[t!]
	\tablefontsize
    \centerline{
\begin{tabular}{l>{\raggedright\arraybackslash}p{3.5cm}>{\raggedright\arraybackslash}p{3.5cm}>{\raggedright\arraybackslash}p{3.5cm}}
\toprule
Name & Lit-BM & HuRI & HI-union \\
Kind & Undirected, unweighted & Undirected, unweighted & Undirected, unweighted \\
Nodes are & Human proteins & Human proteins & Human proteins \\
Links are & Binary protein interactions & Binary protein interactions & Binary protein interactions \\
Considerations &  &  & HI-union includes HuRI \\
\midrule
Number of nodes & 6\,047 & 8\,272 & 9\,094 \\
Number of links & 13\,441 [683 self-loops] & 52\,548 [480 self-loops] & 64\,006 [764 self-loops] \\
Degree$^*$ & 4.446 [1, 415] & 12.705 [1, 500] & 14.077 [1, 641] \\
Clustering & 0.0618 & 0.0592 & 0.0621 \\
Connected & 248 components [92.31\% in largest] & 72 components [98.51\% in largest] & 70 components [98.81\% in largest] \\
Component size$^*$ & 24.383 [1, 5\,582] & 114.889 [1, 8\,149] & 129.914 [1, 8\,986] \\
Diameter & n/a & n/a & n/a \\
Largest component's diameter & 13 & 12 & 11 \\
Assortativity (degree) & -0.0876 & -0.115 & -0.131 \\
\midrule
Node metadata & GENCODE v27 gene annotations & GENCODE v27 gene annotations & GENCODE v27 gene annotations \\
Link metadata & None & None & None \\
Date of creation & 2019 & 2019 & 2005--2019 \\
Data generating process & Links taken from literature-curated data set & Links inferred using a high-throughput three-panel yeast two-hybrid assay applied to pairs of protein-encoding genes taken from human ORFeome v9.1 & Links taken from union of previous PPI screens \\
Funding & National Institutes of Health and others & National Institutes of Health and others & National Institutes of Health and others \\
Citation & Luck \emph{et al.}\ (2020) \cite{luck2020} & Luck \emph{et al.}\ (2020) \cite{luck2020} & Luck \emph{et al.}\ (2020) \cite{luck2020} \\
Access & \multicolumn{3}{l}{\url{http://www.interactome-atlas.org/download} (accessed 2022-04-01)} \\
\bottomrule
\multicolumn{2}{l}{\footnotesize $^*$Distributions summarized with average [min, max].}
\end{tabular}
}
%
%
%
%
    \caption{A network ``multicard''. Here the HuRI network (\cref{tab:huri}) is compared against two other networks from the same study \cite{luck2020}. A reader can quickly ascertain similarities and differences between the networks.
    \label{tab:huri-multicard}}
\end{table}

\section{Discussion}
\label{sec:discussion}

In this paper we propose network cards, simple and accessible tabular summaries of network datasets. 
Network cards are intended to be concise, readable, and flexible.
Using a corpus of example networks, we highlight the information contained within network cards and how researchers can employ them in their own work.
To help researchers use network cards, we have created a schema, fill-in templates, and an open source software package for generating cards, all available at \reponame{}.

We envision network cards being useful in the following situations:
(1) as tables in manuscripts and supporting material;
(2) as summaries display on pages of online repositories of network data;
(3) included with data downloads as metadata alongside ``READMEs'' and other information;
(4) as reporting guidelines or checklists adopted specifically for studies using network data;
(5) shown as part of internal presentations with collaborators working on shared data;
(6) lastly, and with the caveat that dense technical information may not be appropriate for some venues,
with broad adoption, network cards may also be useful in presentations during conferences and meetings.

Broad adoption of network cards may lead to three potential benefits.
One, it becomes easier to understand papers using network data. 
Readers can more quickly grasp the most salient details of the network or networks employed in the study---what are the nodes, what are the links, when was the network data collected---when those details are presented in a standard manner the reader is accustomed to. 
Quick and accessible information summaries are increasingly important as the volume of scientific research grows~\cite{kostoffStructuredAbstracts2001}.

A second potential benefit of network cards stems from their highlighting of non-structural information such as ethical concerns or the presence of metadata, serving as a useful checklist. 
By drawing attention to these facets of the data, readers are better equipped to understand the appropriateness and broader consequences of working with the network data, especially important for data that come with ethical or privacy concerns.
Highlighting these details is important if the data are available and the reader wishes to use it themselves~\cite{gebru2021datasheets}. 
Those details, which may be lost when considering only the network structure, may reveal that the data may not be suitable for certain purposes.
The succinctness of network cards can increase the chances for researchers to correctly identify which data to use for themselves.

Lastly, a third benefit of broad use of network cards comes through automation.
It is more common for papers to examine a corpus of hundreds or even thousands of different networks~\cite{kunegis2013konect,rossi2015network,kujala2018a,Broido_Clauset_2019,voitalov2019scale,lynn2020human}. 
Analyzing networks at this scale is invaluable for revealing broad patterns and trends across research domains~\cite{ikehara2017characterizing}.
But such scale requires automation: code must be written to analyze each network programmatically.
Network cards admit a machine-readable JSON format, for which we provide a schema.
If network cards are created when networks are added to large corpora, then subsequent analysis programs can read those cards at the same time they read the network data itself.
In other words, standardizing the representation of network meta-information using cards has the potential to make that meta-information computationally accessible %
which can then drive a deeper understanding of network corpora.

\appendix

\section{Card contents}
\label{sec:cardcontents}

Here we list all the entries that constitute the standard three-panel network card.
These entries are organized by panel and the descriptions were derived from a schema we have drafted to help with the standardization process. 

{\small
\begin{description}\singlespacing
\item[Overall]
    Describes the network type (directed, weighted, etc.), what do nodes and links represent, what other key considerations do the data entail.

\begin{description}
\item[Name] A written identifier for the network
\item[Kind] Is the network undirected, unweighted, etc..
\item[Nodes are] The definition of nodes.
\item[Links are] The definition of links (edges).
\item[Link weights are] The definition of link (edge) weights.
\item[Considerations] What considerations should be taken into account regarding the network's overall properties.
\end{description}

\item[Structure]
    Summary statistics for the size, density, and other properties of the network structure.

\begin{description}
\item[Number of nodes] The number of nodes in the network.
\item[Number of links] The number of links (edges) in the network.
\item[Degree] A summary of the degree distribution.
\item[Clustering] The average clustering of the network.
\item[Connected] A description and summary of the network's connectivity.
\item[Component size] An optional description about the sizes of the network's components.
\item[Diameter] The diameter of the network, if connected.
\item[Largest component's diameter] The diameter of the largest component's induced subgraph, if network is not connected.
\item[Assortativity (degree)] The degree assortativity of the network.
\end{description}

\item[Meta-information]
    Further details such as the presence of any node or link metadata, data collection documentation, ethical considerations, citations, and any funding acknowledgments.

\begin{description}
\item[Node metadata] A description of any metadata associated with nodes.
\item[Link metadata] A description of any metadata associated with links.
\item[Date of creation] A description of when the network data were gathered or created.
\item[Data generating process] A description of how the network data were generated.
\item[Ethics] A description of ethical considerations for the network data.
\item[Funding] A description of funding related to the network data.
\item[Citation] A citation and/or DOI associated with the network data.
\item[Access] A URL or other description of where data can be obtained.\end{description}
\end{description} }

For more details, please see \reponame{}.

\begin{backmatter}

\section*{Funding}%
J.B.\ acknowledges support by Google Open Source under the Open Source Complex Ecosystems And Networks (OCEAN) project.
Y.-Y.A.\ acknowledges support by the Air Force Office of Scientific Research under award number FA9550-19-1-0391.

\section*{Availability of data and materials}%

Network data for \emph{Zachary Karate Club} (\cref{tab:zachary_karate_club} \cite{zachary1977}) was retrieved from \url{https://nrvis.com/download/data/soc/soc-karate.zip} on 12 February 2022.
Network data for \emph{M\_PL\_058} (\cref{tab:M_PL_058} \cite{bartomeus2008}) was retrieved from \url{https://www.web-of-life.es/networkjson.php?id=M_PL_058} on 10 March 2022.
Network data for \emph{TNet Malawi Pilot} (\cref{tab:tnet_malawi_pilot} \cite{ozella2021}) was retrieved from \url{http://www.sociopatterns.org/datasets/contact-patterns-in-a-village-in-rural-malawi/} on 24 March 2022.
Network data for \emph{OpenFlights routes} (\cref{tab:openflights_directed}) was retrieved from \url{https://openflights.org/data.html} on 26 September 2022.
Network data for \emph{HuRI}, \emph{Lit-BM}, and \emph{HI-union} (\cref{tab:huri,tab:huri-multicard} \cite{luck2020}) were retrieved from
\url{http://www.interactome-atlas.org/download} on 1 April 2022.
Data files as used in the study have been deposited at \url{https://doi.org/10.6084/m9.figshare.20286648}.

\section*{Competing interests}
The authors declare that they have no competing interests.

\section*{Authors' contributions}
J.B. and Y.-Y.A. designed and conducted the research and wrote the manuscript.

%
%

%
%
%
%
%
%
%
%
%
%
%
%
%
%

%

\newcommand{\BMCxmlcomment}[1]{}

\BMCxmlcomment{

<refgrp>

<bibl id="B1">
  <title><p>Network science</p></title>
  <aug>
    <au><snm>B{\"o}rner</snm><fnm>K</fnm></au>
    <au><snm>Sanyal</snm><fnm>S</fnm></au>
    <au><snm>Vespignani</snm><fnm>A</fnm></au>
  </aug>
  <source>Annual Review of Information Science and Technology</source>
  <pubdate>2007</pubdate>
  <volume>41</volume>
  <issue>1</issue>
  <fpage>537</fpage>
  <lpage>-607</lpage>
</bibl>

<bibl id="B2">
  <title><p>Complexity: a guided tour</p></title>
  <aug>
    <au><snm>Mitchell</snm><fnm>M</fnm></au>
  </aug>
  <publisher>Oxford [England]; New York: Oxford University Press</publisher>
  <pubdate>2009</pubdate>
</bibl>

<bibl id="B3">
  <title><p>Networks: An Introduction</p></title>
  <aug>
    <au><snm>Newman</snm><fnm>M. E. J</fnm></au>
  </aug>
  <publisher>Oxford: Oxford University Press</publisher>
  <edition>2</edition>
  <pubdate>2018</pubdate>
</bibl>

<bibl id="B4">
  <title><p>A First Course in Network Science</p></title>
  <aug>
    <au><snm>Menczer</snm><fnm>F</fnm></au>
    <au><snm>Fortunato</snm><fnm>S</fnm></au>
    <au><snm>Davis</snm><fnm>CA</fnm></au>
  </aug>
  <publisher>Cambridge: Cambridge University Press</publisher>
  <pubdate>2020</pubdate>
</bibl>

<bibl id="B5">
  <title><p>Computational Social Science</p></title>
  <aug>
    <au><snm>Lazer</snm><fnm>D</fnm></au>
    <au><snm>Pentland</snm><fnm>A</fnm></au>
    <au><snm>Adamic</snm><fnm>L</fnm></au>
    <au><snm>Aral</snm><fnm>S</fnm></au>
    <au><snm>Barabási</snm><fnm>AL</fnm></au>
    <au><snm>Brewer</snm><fnm>D</fnm></au>
    <au><snm>Christakis</snm><fnm>N</fnm></au>
    <au><snm>Contractor</snm><fnm>N</fnm></au>
    <au><snm>Fowler</snm><fnm>J</fnm></au>
    <au><snm>Gutmann</snm><fnm>M</fnm></au>
    <au><snm>Jebara</snm><fnm>T</fnm></au>
    <au><snm>King</snm><fnm>G</fnm></au>
    <au><snm>Macy</snm><fnm>M</fnm></au>
    <au><snm>Roy</snm><fnm>D</fnm></au>
    <au><snm>Van Alstyne</snm><fnm>M</fnm></au>
  </aug>
  <source>Science</source>
  <pubdate>2009</pubdate>
  <volume>323</volume>
  <issue>5915</issue>
  <fpage>721–723</fpage>
</bibl>

<bibl id="B6">
  <title><p>Model Cards for Model Reporting</p></title>
  <aug>
    <au><snm>Mitchell</snm><fnm>M</fnm></au>
    <au><snm>Wu</snm><fnm>S</fnm></au>
    <au><snm>Zaldivar</snm><fnm>A</fnm></au>
    <au><snm>Barnes</snm><fnm>P</fnm></au>
    <au><snm>Vasserman</snm><fnm>L</fnm></au>
    <au><snm>Hutchinson</snm><fnm>B</fnm></au>
    <au><snm>Spitzer</snm><fnm>E</fnm></au>
    <au><snm>Raji</snm><fnm>ID</fnm></au>
    <au><snm>Gebru</snm><fnm>T</fnm></au>
  </aug>
  <source>Proceedings of the Conference on Fairness, Accountability, and
  Transparency</source>
  <publisher>Atlanta GA USA: ACM</publisher>
  <pubdate>2019</pubdate>
  <fpage>220–229</fpage>
</bibl>

<bibl id="B7">
  <title><p>Datasheets for datasets</p></title>
  <aug>
    <au><snm>Gebru</snm><fnm>T</fnm></au>
    <au><snm>Morgenstern</snm><fnm>J</fnm></au>
    <au><snm>Vecchione</snm><fnm>B</fnm></au>
    <au><snm>Vaughan</snm><fnm>JW</fnm></au>
    <au><snm>Wallach</snm><fnm>H</fnm></au>
    <au><snm>{Daumé III}</snm><fnm>H</fnm></au>
    <au><snm>Crawford</snm><fnm>K</fnm></au>
  </aug>
  <source>Communications of the ACM</source>
  <pubdate>2021</pubdate>
  <volume>64</volume>
  <issue>12</issue>
  <fpage>86–92</fpage>
</bibl>

<bibl id="B8">
  <title><p>An information flow model for conflict and fission in small
  groups</p></title>
  <aug>
    <au><snm>Zachary</snm><fnm>WW</fnm></au>
  </aug>
  <source>Journal of anthropological research</source>
  <publisher>University of New Mexico</publisher>
  <pubdate>1977</pubdate>
  <volume>33</volume>
  <issue>4</issue>
  <fpage>452</fpage>
  <lpage>-473</lpage>
</bibl>

<bibl id="B9">
  <title><p>Yeast Two-Hybrid, a Powerful Tool for Systems Biology</p></title>
  <aug>
    <au><snm>Brückner</snm><fnm>A</fnm></au>
    <au><snm>Polge</snm><fnm>C</fnm></au>
    <au><snm>Lentze</snm><fnm>N</fnm></au>
    <au><snm>Auerbach</snm><fnm>D</fnm></au>
    <au><snm>Schlattner</snm><fnm>U</fnm></au>
  </aug>
  <source>International Journal of Molecular Sciences</source>
  <pubdate>2009</pubdate>
  <volume>10</volume>
  <issue>6</issue>
  <fpage>2763–2788</fpage>
</bibl>

<bibl id="B10">
  <title><p>Analysis of protein complexes using mass spectrometry</p></title>
  <aug>
    <au><snm>Gingras</snm><fnm>AC</fnm></au>
    <au><snm>Gstaiger</snm><fnm>M</fnm></au>
    <au><snm>Raught</snm><fnm>B</fnm></au>
    <au><snm>Aebersold</snm><fnm>R</fnm></au>
  </aug>
  <source>Nature Reviews Molecular Cell Biology</source>
  <pubdate>2007</pubdate>
  <volume>8</volume>
  <issue>8</issue>
  <fpage>645–654</fpage>
</bibl>

<bibl id="B11">
  <title><p>Growth rates of modern science: A bibliometric analysis based on
  the number of publications and cited references: Growth Rates of Modern
  Science: A Bibliometric Analysis Based on the Number of Publications and
  Cited References</p></title>
  <aug>
    <au><snm>Bornmann</snm><fnm>L</fnm></au>
    <au><snm>Mutz</snm><fnm>R</fnm></au>
  </aug>
  <source>Journal of the Association for Information Science and
  Technology</source>
  <pubdate>2015</pubdate>
  <volume>66</volume>
  <issue>11</issue>
  <fpage>2215–2222</fpage>
</bibl>

<bibl id="B12">
  <title><p>Science of science</p></title>
  <aug>
    <au><snm>Fortunato</snm><fnm>S</fnm></au>
    <au><snm>Bergstrom</snm><fnm>CT</fnm></au>
    <au><snm>Börner</snm><fnm>K</fnm></au>
    <au><snm>Evans</snm><fnm>JA</fnm></au>
    <au><snm>Helbing</snm><fnm>D</fnm></au>
    <au><snm>Milojević</snm><fnm>S</fnm></au>
    <au><snm>Petersen</snm><fnm>AM</fnm></au>
    <au><snm>Radicchi</snm><fnm>F</fnm></au>
    <au><snm>Sinatra</snm><fnm>R</fnm></au>
    <au><snm>Uzzi</snm><fnm>B</fnm></au>
    <au><snm>Vespignani</snm><fnm>A</fnm></au>
    <au><snm>Waltman</snm><fnm>L</fnm></au>
    <au><snm>Wang</snm><fnm>D</fnm></au>
    <au><snm>Barabási</snm><fnm>AL</fnm></au>
  </aug>
  <source>Science</source>
  <pubdate>2018</pubdate>
  <volume>359</volume>
  <issue>6379</issue>
  <fpage>eaao0185</fpage>
</bibl>

<bibl id="B13">
  <title><p>Growth rates of modern science: a latent piecewise growth curve
  approach to model publication numbers from established and new literature
  databases</p></title>
  <aug>
    <au><snm>Bornmann</snm><fnm>L</fnm></au>
    <au><snm>Haunschild</snm><fnm>R</fnm></au>
    <au><snm>Mutz</snm><fnm>R</fnm></au>
  </aug>
  <source>Humanities and Social Sciences Communications</source>
  <pubdate>2021</pubdate>
  <volume>8</volume>
  <issue>1</issue>
  <fpage>224</fpage>
</bibl>

<bibl id="B14">
  <title><p>Why Most Published Research Findings Are False</p></title>
  <aug>
    <au><snm>Ioannidis</snm><fnm>JPA</fnm></au>
  </aug>
  <source>PLoS Medicine</source>
  <pubdate>2005</pubdate>
  <volume>2</volume>
  <issue>8</issue>
  <fpage>e124</fpage>
</bibl>

<bibl id="B15">
  <title><p>Estimating the reproducibility of psychological science</p></title>
  <aug>
    <au><snm>Collaboration</snm><fnm>OS</fnm></au>
  </aug>
  <source>Science</source>
  <pubdate>2015</pubdate>
  <volume>349</volume>
  <issue>6251</issue>
  <fpage>aac4716</fpage>
</bibl>

<bibl id="B16">
  <title><p>Publication bias and the canonization of false facts</p></title>
  <aug>
    <au><snm>Nissen</snm><fnm>SB</fnm></au>
    <au><snm>Magidson</snm><fnm>T</fnm></au>
    <au><snm>Gross</snm><fnm>K</fnm></au>
    <au><snm>Bergstrom</snm><fnm>CT</fnm></au>
  </aug>
  <source>eLife</source>
  <pubdate>2016</pubdate>
  <volume>5</volume>
  <fpage>e21451</fpage>
</bibl>

<bibl id="B17">
  <title><p>Threats of a replication crisis in empirical computer
  science</p></title>
  <aug>
    <au><snm>Cockburn</snm><fnm>A</fnm></au>
    <au><snm>Dragicevic</snm><fnm>P</fnm></au>
    <au><snm>Besançon</snm><fnm>L</fnm></au>
    <au><snm>Gutwin</snm><fnm>C</fnm></au>
  </aug>
  <source>Communications of the ACM</source>
  <pubdate>2020</pubdate>
  <volume>63</volume>
  <issue>8</issue>
  <fpage>70–79</fpage>
</bibl>

<bibl id="B18">
  <title><p>Measurement error and the replication crisis</p></title>
  <aug>
    <au><snm>Loken</snm><fnm>E</fnm></au>
    <au><snm>Gelman</snm><fnm>A</fnm></au>
  </aug>
  <source>Science</source>
  <pubdate>2017</pubdate>
  <volume>355</volume>
  <issue>6325</issue>
  <fpage>584–585</fpage>
</bibl>

<bibl id="B19">
  <title><p>Redefine statistical significance</p></title>
  <aug>
    <au><snm>Benjamin</snm><fnm>DJ</fnm></au>
    <au><snm>Berger</snm><fnm>JO</fnm></au>
    <au><snm>Johannesson</snm><fnm>M</fnm></au>
    <au><snm>Nosek</snm><fnm>BA</fnm></au>
    <au><snm>Wagenmakers</snm><fnm>E. J.</fnm></au>
    <au><snm>Berk</snm><fnm>R</fnm></au>
    <au><snm>Bollen</snm><fnm>KA</fnm></au>
    <au><snm>Brembs</snm><fnm>B</fnm></au>
    <au><snm>Brown</snm><fnm>L</fnm></au>
    <au><snm>Camerer</snm><fnm>C</fnm></au>
    <au><snm>Cesarini</snm><fnm>D</fnm></au>
    <au><snm>Chambers</snm><fnm>CD</fnm></au>
    <au><snm>Clyde</snm><fnm>M</fnm></au>
    <au><snm>Cook</snm><fnm>TD</fnm></au>
    <au><snm>De Boeck</snm><fnm>P</fnm></au>
    <au><snm>Dienes</snm><fnm>Z</fnm></au>
    <au><snm>Dreber</snm><fnm>A</fnm></au>
    <au><snm>Easwaran</snm><fnm>K</fnm></au>
    <au><snm>Efferson</snm><fnm>C</fnm></au>
    <au><snm>Fehr</snm><fnm>E</fnm></au>
    <au><snm>Fidler</snm><fnm>F</fnm></au>
    <au><snm>Field</snm><fnm>AP</fnm></au>
    <au><snm>Forster</snm><fnm>M</fnm></au>
    <au><snm>George</snm><fnm>EI</fnm></au>
    <au><snm>Gonzalez</snm><fnm>R</fnm></au>
    <au><snm>Goodman</snm><fnm>S</fnm></au>
    <au><snm>Green</snm><fnm>E</fnm></au>
    <au><snm>Green</snm><fnm>DP</fnm></au>
    <au><snm>Greenwald</snm><fnm>AG</fnm></au>
    <au><snm>Hadfield</snm><fnm>JD</fnm></au>
    <au><snm>Hedges</snm><fnm>LV</fnm></au>
    <au><snm>Held</snm><fnm>L</fnm></au>
    <au><snm>Hua Ho</snm><fnm>T</fnm></au>
    <au><snm>Hoijtink</snm><fnm>H</fnm></au>
    <au><snm>Hruschka</snm><fnm>DJ</fnm></au>
    <au><snm>Imai</snm><fnm>K</fnm></au>
    <au><snm>Imbens</snm><fnm>G</fnm></au>
    <au><snm>Ioannidis</snm><fnm>JPA</fnm></au>
    <au><snm>Jeon</snm><fnm>M</fnm></au>
    <au><snm>Jones</snm><fnm>JH</fnm></au>
    <au><snm>Kirchler</snm><fnm>M</fnm></au>
    <au><snm>Laibson</snm><fnm>D</fnm></au>
    <au><snm>List</snm><fnm>J</fnm></au>
    <au><snm>Little</snm><fnm>R</fnm></au>
    <au><snm>Lupia</snm><fnm>A</fnm></au>
    <au><snm>Machery</snm><fnm>E</fnm></au>
    <au><snm>Maxwell</snm><fnm>SE</fnm></au>
    <au><snm>McCarthy</snm><fnm>M</fnm></au>
    <au><snm>Moore</snm><fnm>DA</fnm></au>
    <au><snm>Morgan</snm><fnm>SL</fnm></au>
    <au><snm>Munafó</snm><fnm>M</fnm></au>
    <au><snm>Nakagawa</snm><fnm>S</fnm></au>
    <au><snm>Nyhan</snm><fnm>B</fnm></au>
    <au><snm>Parker</snm><fnm>TH</fnm></au>
    <au><snm>Pericchi</snm><fnm>L</fnm></au>
    <au><snm>Perugini</snm><fnm>M</fnm></au>
    <au><snm>Rouder</snm><fnm>J</fnm></au>
    <au><snm>Rousseau</snm><fnm>J</fnm></au>
    <au><snm>Savalei</snm><fnm>V</fnm></au>
    <au><snm>Schönbrodt</snm><fnm>FD</fnm></au>
    <au><snm>Sellke</snm><fnm>T</fnm></au>
    <au><snm>Sinclair</snm><fnm>B</fnm></au>
    <au><snm>Tingley</snm><fnm>D</fnm></au>
    <au><snm>Van Zandt</snm><fnm>T</fnm></au>
    <au><snm>Vazire</snm><fnm>S</fnm></au>
    <au><snm>Watts</snm><fnm>DJ</fnm></au>
    <au><snm>Winship</snm><fnm>C</fnm></au>
    <au><snm>Wolpert</snm><fnm>RL</fnm></au>
    <au><snm>Xie</snm><fnm>Y</fnm></au>
    <au><snm>Young</snm><fnm>C</fnm></au>
    <au><snm>Zinman</snm><fnm>J</fnm></au>
    <au><snm>Johnson</snm><fnm>VE</fnm></au>
  </aug>
  <source>Nature Human Behaviour</source>
  <pubdate>2018</pubdate>
  <volume>2</volume>
  <issue>1</issue>
  <fpage>6–10</fpage>
</bibl>

<bibl id="B20">
  <title><p>Statistical Analysis Must Improve to Address the Reproducibility
  Crisis: The ACcess to Transparent Statistics (ACTS) Call to
  Action</p></title>
  <aug>
    <au><snm>Gosselin</snm><fnm>R</fnm></au>
  </aug>
  <source>BioEssays</source>
  <pubdate>2020</pubdate>
  <volume>42</volume>
  <issue>1</issue>
  <fpage>1900189</fpage>
</bibl>

<bibl id="B21">
  <title><p>Investigating reproducibility and tracking provenance – A genomic
  workflow case study</p></title>
  <aug>
    <au><snm>Kanwal</snm><fnm>S</fnm></au>
    <au><snm>Khan</snm><fnm>FZ</fnm></au>
    <au><snm>Lonie</snm><fnm>A</fnm></au>
    <au><snm>Sinnott</snm><fnm>RO</fnm></au>
  </aug>
  <source>BMC Bioinformatics</source>
  <pubdate>2017</pubdate>
  <volume>18</volume>
  <issue>1</issue>
  <fpage>337</fpage>
</bibl>

<bibl id="B22">
  <title><p>CRISIS, WHAT CRISIS – DOES REPRODUCIBILITY IN MODELING &amp
  SIMULATION REALLY MATTER?</p></title>
  <aug>
    <au><snm>Taylor</snm><fnm>SJE</fnm></au>
    <au><snm>Eldabi</snm><fnm>T</fnm></au>
    <au><snm>Monks</snm><fnm>T</fnm></au>
    <au><snm>Rabe</snm><fnm>M</fnm></au>
    <au><snm>Uhrmacher</snm><fnm>AM</fnm></au>
  </aug>
  <source>2018 Winter Simulation Conference (WSC)</source>
  <publisher>Gothenburg, Sweden: IEEE</publisher>
  <pubdate>2018</pubdate>
  <fpage>749–762</fpage>
</bibl>

<bibl id="B23">
  <title><p>Improving reproducibility of data science pipelines through
  transparent provenance capture</p></title>
  <aug>
    <au><snm>Rupprecht</snm><fnm>L</fnm></au>
    <au><snm>Davis</snm><fnm>JC</fnm></au>
    <au><snm>Arnold</snm><fnm>C</fnm></au>
    <au><snm>Gur</snm><fnm>Y</fnm></au>
    <au><snm>Bhagwat</snm><fnm>D</fnm></au>
  </aug>
  <source>Proceedings of the VLDB Endowment</source>
  <pubdate>2020</pubdate>
  <volume>13</volume>
  <issue>12</issue>
  <fpage>3354–3368</fpage>
</bibl>

<bibl id="B24">
  <title><p>Contrasting effects of invasive plants in plant–pollinator
  networks</p></title>
  <aug>
    <au><snm>Bartomeus</snm><fnm>I</fnm></au>
    <au><snm>Vilà</snm><fnm>M</fnm></au>
    <au><snm>Santamaría</snm><fnm>L</fnm></au>
  </aug>
  <source>Oecologia</source>
  <pubdate>2008</pubdate>
  <volume>155</volume>
  <issue>4</issue>
  <fpage>761–770</fpage>
</bibl>

<bibl id="B25">
  <title><p>Using wearable proximity sensors to characterize social contact
  patterns in a village of rural Malawi</p></title>
  <aug>
    <au><snm>Ozella</snm><fnm>L</fnm></au>
    <au><snm>Paolotti</snm><fnm>D</fnm></au>
    <au><snm>Lichand</snm><fnm>G</fnm></au>
    <au><snm>Rodr{\'\i}guez</snm><fnm>JP</fnm></au>
    <au><snm>Haenni</snm><fnm>S</fnm></au>
    <au><snm>Phuka</snm><fnm>J</fnm></au>
    <au><snm>Leal Neto</snm><fnm>OB</fnm></au>
    <au><snm>Cattuto</snm><fnm>C</fnm></au>
  </aug>
  <source>EPJ Data Science</source>
  <pubdate>2021</pubdate>
  <volume>10</volume>
  <issue>1</issue>
  <fpage>46</fpage>
</bibl>

<bibl id="B26">
  <title><p>A reference map of the human binary protein interactome</p></title>
  <aug>
    <au><snm>Luck</snm><fnm>K</fnm></au>
    <au><snm>Kim</snm><fnm>DK</fnm></au>
    <au><snm>Lambourne</snm><fnm>L</fnm></au>
    <au><snm>Spirohn</snm><fnm>K</fnm></au>
    <au><snm>Begg</snm><fnm>BE</fnm></au>
    <au><snm>Bian</snm><fnm>W</fnm></au>
    <au><snm>Brignall</snm><fnm>R</fnm></au>
    <au><snm>Cafarelli</snm><fnm>T</fnm></au>
    <au><snm>Campos Laborie</snm><fnm>FJ</fnm></au>
    <au><snm>Charloteaux</snm><fnm>B</fnm></au>
    <au><snm>Choi</snm><fnm>D</fnm></au>
    <au><snm>Coté</snm><fnm>AG</fnm></au>
    <au><snm>Daley</snm><fnm>M</fnm></au>
    <au><snm>Deimling</snm><fnm>S</fnm></au>
    <au><snm>Desbuleux</snm><fnm>A</fnm></au>
    <au><snm>Dricot</snm><fnm>A</fnm></au>
    <au><snm>Gebbia</snm><fnm>M</fnm></au>
    <au><snm>Hardy</snm><fnm>MF</fnm></au>
    <au><snm>Kishore</snm><fnm>N</fnm></au>
    <au><snm>Knapp</snm><fnm>JJ</fnm></au>
    <au><snm>Kovács</snm><fnm>IA</fnm></au>
    <au><snm>Lemmens</snm><fnm>I</fnm></au>
    <au><snm>Mee</snm><fnm>MW</fnm></au>
    <au><snm>Mellor</snm><fnm>JC</fnm></au>
    <au><snm>Pollis</snm><fnm>C</fnm></au>
    <au><snm>Pons</snm><fnm>C</fnm></au>
    <au><snm>Richardson</snm><fnm>AD</fnm></au>
    <au><snm>Schlabach</snm><fnm>S</fnm></au>
    <au><snm>Teeking</snm><fnm>B</fnm></au>
    <au><snm>Yadav</snm><fnm>A</fnm></au>
    <au><snm>Babor</snm><fnm>M</fnm></au>
    <au><snm>Balcha</snm><fnm>D</fnm></au>
    <au><snm>Basha</snm><fnm>O</fnm></au>
    <au><snm>Bowman Colin</snm><fnm>C</fnm></au>
    <au><snm>Chin</snm><fnm>SF</fnm></au>
    <au><snm>Choi</snm><fnm>SG</fnm></au>
    <au><snm>Colabella</snm><fnm>C</fnm></au>
    <au><snm>Coppin</snm><fnm>G</fnm></au>
    <au><snm>D’Amata</snm><fnm>C</fnm></au>
    <au><snm>De Ridder</snm><fnm>D</fnm></au>
    <au><snm>De Rouck</snm><fnm>S</fnm></au>
    <au><snm>Duran Frigola</snm><fnm>M</fnm></au>
    <au><snm>Ennajdaoui</snm><fnm>H</fnm></au>
    <au><snm>Goebels</snm><fnm>F</fnm></au>
    <au><snm>Goehring</snm><fnm>L</fnm></au>
    <au><snm>Gopal</snm><fnm>A</fnm></au>
    <au><snm>Haddad</snm><fnm>G</fnm></au>
    <au><snm>Hatchi</snm><fnm>E</fnm></au>
    <au><snm>Helmy</snm><fnm>M</fnm></au>
    <au><snm>Jacob</snm><fnm>Y</fnm></au>
    <au><snm>Kassa</snm><fnm>Y</fnm></au>
    <au><snm>Landini</snm><fnm>S</fnm></au>
    <au><snm>Li</snm><fnm>R</fnm></au>
    <au><snm>Lieshout</snm><fnm>N</fnm></au>
    <au><snm>MacWilliams</snm><fnm>A</fnm></au>
    <au><snm>Markey</snm><fnm>D</fnm></au>
    <au><snm>Paulson</snm><fnm>JN</fnm></au>
    <au><snm>Rangarajan</snm><fnm>S</fnm></au>
    <au><snm>Rasla</snm><fnm>J</fnm></au>
    <au><snm>Rayhan</snm><fnm>A</fnm></au>
    <au><snm>Rolland</snm><fnm>T</fnm></au>
    <au><snm>San Miguel</snm><fnm>A</fnm></au>
    <au><snm>Shen</snm><fnm>Y</fnm></au>
    <au><snm>Sheykhkarimli</snm><fnm>D</fnm></au>
    <au><snm>Sheynkman</snm><fnm>GM</fnm></au>
    <au><snm>Simonovsky</snm><fnm>E</fnm></au>
    <au><snm>Taşan</snm><fnm>M</fnm></au>
    <au><snm>Tejeda</snm><fnm>A</fnm></au>
    <au><snm>Tropepe</snm><fnm>V</fnm></au>
    <au><snm>Twizere</snm><fnm>JC</fnm></au>
    <au><snm>Wang</snm><fnm>Y</fnm></au>
    <au><snm>Weatheritt</snm><fnm>RJ</fnm></au>
    <au><snm>Weile</snm><fnm>J</fnm></au>
    <au><snm>Xia</snm><fnm>Y</fnm></au>
    <au><snm>Yang</snm><fnm>X</fnm></au>
    <au><snm>Yeger Lotem</snm><fnm>E</fnm></au>
    <au><snm>Zhong</snm><fnm>Q</fnm></au>
    <au><snm>Aloy</snm><fnm>P</fnm></au>
    <au><snm>Bader</snm><fnm>GD</fnm></au>
    <au><snm>De Las Rivas</snm><fnm>J</fnm></au>
    <au><snm>Gaudet</snm><fnm>S</fnm></au>
    <au><snm>Hao</snm><fnm>T</fnm></au>
    <au><snm>Rak</snm><fnm>J</fnm></au>
    <au><snm>Tavernier</snm><fnm>J</fnm></au>
    <au><snm>Hill</snm><fnm>DE</fnm></au>
    <au><snm>Vidal</snm><fnm>M</fnm></au>
    <au><snm>Roth</snm><fnm>FP</fnm></au>
    <au><snm>Calderwood</snm><fnm>MA</fnm></au>
  </aug>
  <source>Nature</source>
  <pubdate>2020</pubdate>
  <volume>580</volume>
  <issue>7803</issue>
  <fpage>402–408</fpage>
</bibl>

<bibl id="B27">
  <title><p>Structured Abstracts for Technical Journals</p></title>
  <aug>
    <au><snm>Kostoff</snm><fnm>RN</fnm></au>
    <au><snm>Hartley</snm><fnm>J</fnm></au>
  </aug>
  <source>Science</source>
  <pubdate>2001</pubdate>
  <volume>292</volume>
  <issue>5519</issue>
  <fpage>1067</fpage>
  <lpage>1067</lpage>
</bibl>

<bibl id="B28">
  <title><p>{KONECT}: the {Koblenz} network collection</p></title>
  <aug>
    <au><snm>Kunegis</snm><fnm>J</fnm></au>
  </aug>
  <source>Proceedings of the 22nd International Conference on World Wide
  Web</source>
  <publisher>Rio de Janeiro Brazil: ACM</publisher>
  <pubdate>2013</pubdate>
  <fpage>1343–1350</fpage>
</bibl>

<bibl id="B29">
  <title><p>The Network Data Repository with Interactive Graph Analytics and
  Visualization</p></title>
  <aug>
    <au><snm>Rossi</snm><fnm>RA</fnm></au>
    <au><snm>Ahmed</snm><fnm>NK</fnm></au>
  </aug>
  <source>Proceedings of the Twenty-Ninth AAAI Conference on Artificial
  Intelligence</source>
  <series><title><p>AAAI'15</p></title></series>
  <pubdate>2015</pubdate>
  <fpage>4292–4293</fpage>
</bibl>

<bibl id="B30">
  <title><p>A collection of public transport network data sets for 25
  cities</p></title>
  <aug>
    <au><snm>Kujala</snm><fnm>R</fnm></au>
    <au><snm>Weckström</snm><fnm>C</fnm></au>
    <au><snm>Darst</snm><fnm>RK</fnm></au>
    <au><snm>Mladenović</snm><fnm>MN</fnm></au>
    <au><snm>Saramäki</snm><fnm>J</fnm></au>
  </aug>
  <source>Scientific Data</source>
  <pubdate>2018</pubdate>
  <volume>5</volume>
  <issue>1</issue>
  <fpage>180089</fpage>
</bibl>

<bibl id="B31">
  <title><p>Scale-free networks are rare</p></title>
  <aug>
    <au><snm>Broido</snm><fnm>AD</fnm></au>
    <au><snm>Clauset</snm><fnm>A</fnm></au>
  </aug>
  <source>Nature Communications</source>
  <pubdate>2019</pubdate>
  <volume>10</volume>
  <issue>1</issue>
  <fpage>1017</fpage>
</bibl>

<bibl id="B32">
  <title><p>Scale-free networks well done</p></title>
  <aug>
    <au><snm>Voitalov</snm><fnm>I</fnm></au>
    <au><snm>Hoorn</snm><fnm>P</fnm></au>
    <au><snm>Hofstad</snm><fnm>R</fnm></au>
    <au><snm>Krioukov</snm><fnm>D</fnm></au>
  </aug>
  <source>Physical Review Research</source>
  <pubdate>2019</pubdate>
  <volume>1</volume>
  <issue>3</issue>
  <fpage>033034</fpage>
</bibl>

<bibl id="B33">
  <title><p>Human information processing in complex networks</p></title>
  <aug>
    <au><snm>Lynn</snm><fnm>CW</fnm></au>
    <au><snm>Papadopoulos</snm><fnm>L</fnm></au>
    <au><snm>Kahn</snm><fnm>AE</fnm></au>
    <au><snm>Bassett</snm><fnm>DS</fnm></au>
  </aug>
  <source>Nature Physics</source>
  <pubdate>2020</pubdate>
  <volume>16</volume>
  <issue>9</issue>
  <fpage>965–973</fpage>
</bibl>

<bibl id="B34">
  <title><p>Characterizing the structural diversity of complex networks across
  domains</p></title>
  <aug>
    <au><snm>Ikehara</snm><fnm>K</fnm></au>
    <au><snm>Clauset</snm><fnm>A</fnm></au>
  </aug>
  <source>arXiv preprint arXiv:1710.11304</source>
  <pubdate>2017</pubdate>
</bibl>

</refgrp>
} 

\end{backmatter}
\end{document}